\documentclass[aps,prb,notitlepage,floatfix,twocolumn,superscriptaddress]{revtex4-2}

\usepackage{amsmath}
\usepackage{amssymb}
\usepackage{mathtools}
\usepackage[tight]{subfigure}
\usepackage{enumitem}
\usepackage{soul}
\usepackage{cancel}
\usepackage{multirow}
\usepackage{tikz}
\usepackage{pifont}
\usepackage{xspace}
\usepackage{comment}
\usepackage{physics}
\usepackage{bbold}
\usepackage{graphicx}
\usepackage{dcolumn}
\usepackage{bm}
\usepackage[breaklinks]{hyperref}
\hypersetup{colorlinks=true, linkcolor=blue, citecolor=blue, filecolor=blue, urlcolor=blue}
\usepackage{ulem}

\begin{document}

\title{Freezing transition in particle-conserving East model}

\author{Cheng Wang}
\affiliation{School of Physics, Peking University, Beijing 100871, China}

\author{Zhi-Cheng Yang}
\email{zcyang19@pku.edu.cn}
\affiliation{School of Physics, Peking University, Beijing 100871, China}
\affiliation{Center for High Energy Physics, Peking University, Beijing 100871, China}

\date{\today}

\begin{abstract}
Quantum kinetically constrained models can exhibit a wealth of dynamical phenomena ranging from anomalous transport to Hilbert-space fragmentation (HSF). We study a class of one-dimensional particle number conserving systems where particle hoppings are subjected to an East-like constraint, akin to facilitated spin models in classical glasses. While such a kinetic constraint leads to HSF, we find that the degree of fragmentation exhibits a sharp transition as the average particle density is varied. Below a critical density, the system transitions from being weakly fragmented where most of the initial states thermalize diffusively, to strongly fragmented where the dynamics are frozen and the system fails to thermalize. 
Remarkably, the East model allows for both efficient numerical simulations and analytic solutions of various diagnostics of the phase transition, from which we obtain a set of exact critical exponents.
We find that the freezing transition in particle-conserving East models belongs to the same universality class as dipole-conserving fracton systems. Our results provide a tractable minimal model for filling-induced freezing transitions associated with HSF, which can be readily tested in state-of-the-art quantum platforms.

\end{abstract}

\maketitle

\textit{Introduction.-} The research field of nonequilibrium quantum many-body dynamics has been a fruitful source of intriguing fundamental questions in theoretical physics over the past few years. While the notion of universality has proved to be a powerful tool in equilibrium statistical mechanics, identifying universality classes in out-of-equilibrium dynamical properties has remained a challenging task. Generic non-integrable quantum many-body systems are expected to thermalize to a maximal entropy state subjected to constraints from conservation laws. Introducing additional ingredients (e.g. disorder, kinetic constraints), however, can impede thermalization and result in a variety of nonequilibrium dynamical phenomena. For example, the Rydberg-blockaded atom array harbors atypical high-energy eigenstates that lead to nonthermal behaviors starting from certain initial states, a phenomenon now known as quantum many-body scars~\cite{turner2018weak, serbyn2021quantum, PhysRevB.98.155134, bernien2017probing}.

\begin{figure}[!t]
\includegraphics[width=0.35\textwidth]{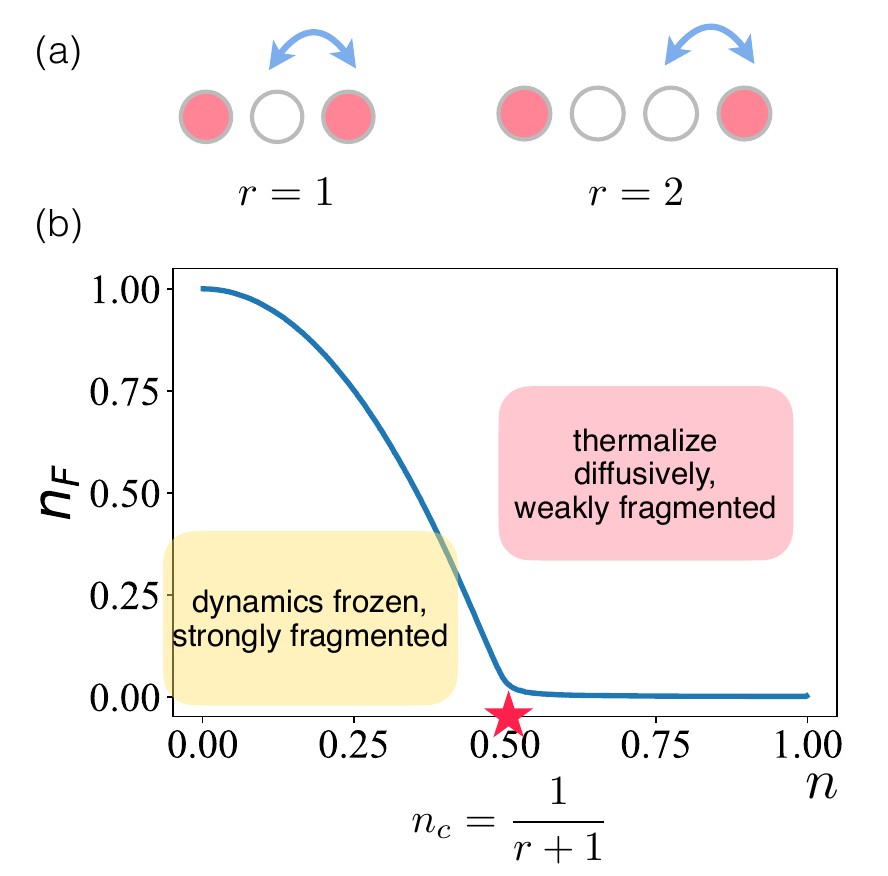}
\caption{(a) Allowed local dynamical moves in the particle-conserving East model with a varying range $r$. A particle is allowed to hop to the right only when there is an occupied site within a distance $r$ to its left. (b) A continuous phase transition at a critical density $n_c=\frac{1}{r+1}$ as diagnosed by the average fraction of frozen sites $\langle n_F \rangle$ (shown for $r=1$). For $n>n_c$, the Hilbert space is strongly fragmented and the dynamics is frozen. Near $n_c$, $\langle n_F \rangle \sim (n_c-n)^\beta$, with $\beta=1$. For $n<n_c$, the Hilbert space is weakly fragmented and charge transport is diffusive. Numerical results are obtained from sampling $10^3$ different configurations of length $L=1000$, for different particle densities. }
\label{fig:east} 
\end{figure}

More generally, one can consider quantum kinetically constrained models, where local dynamical moves are restricted. Such systems are either nonergodic and fail to thermalize~\cite{PhysRevB.92.100305, PhysRevX.10.021051, PhysRevLett.124.207602}, or exhibit anomalously slow relaxation to thermal equilibrium~\cite{PhysRevLett.127.230602, PhysRevB.100.214301, PhysRevResearch.2.033124, PhysRevB.106.L220303}. One paradigmatic example is given by fracton systems~\cite{pretko2020fracton, nandkishore2019fractons}, where particle moves are subjected to both charge (particle number) and dipole moment (center of mass) conservations. It was shown in Refs.~\cite{PhysRevB.101.174204, PhysRevX.10.011047} that the combination of these two conservation laws and locality lead to Hilbert-space fragmentation (HSF): the Hilbert space within a particular symmetry sector further fractures into many disconnected subspaces, giving rise to exponentially many Krylov subsectors in total which cannot be uniquely labelled by the quantum numbers of the conserved charges. One can further quantify the degree of fragmentation and distinguish between \textit{strong} and \textit{weak} fragmentation. Weakly fragmented systems have a dominating Krylov subspace within the symmetry sector, such that typical initial states are able to explore most of the Hilbert space and thermalize. On the other hand, in strongly fragmented systems, an arbitrary initial state is only able to explore a vanishingly small fraction of the entire Hilbert space, and the dynamics is essentially frozen. Interestingly, it was recently demonstrated in Refs.~\cite{PhysRevB.101.214205, PhysRevB.107.045137} that strong and weak fragmentation in fractonic models are separated by a continuous phase transition as the charge density is varied. A natural question that follows is whether such a freezing transition associated with HSF is special to fractonic systems, or does it happen in a broader class of kinetically constrained models. If so, do they belong to the same universality class as fractonic models?

In this Letter, we study a class of one-dimensional systems with a conserved particle number, where particle hoppings are subjected to an East-like constraint, as illustrated in Fig.~\ref{fig:east}(a). As a result of the kinetic constraint, the Hilbert space within a given particle number sector further fractures into Krylov subspaces~\cite{brighi2022hilbert}. We find that the degree of HSF undergoes a sharp transition as the average particle density $n=\frac{N}{L}$ is varied, similarly to fractonic models. While determining the universal scaling properties of the transition in fractonic models has proved to be quite involved~\cite{PhysRevB.101.214205, PhysRevB.107.045137}, the situation is surprisingly simple in East models. We obtain analytic expressions for the critical filling, the size of the largest Krylov subsectors $D_{\rm max}$ for $r=1$, and develop efficient algorithms for computing $D_{\rm max}$ for arbitrary $r$ up to $L \sim 10^3$. We are also able to simulate the exact dynamics of thermal inclusion at infinite times up to $L\sim 10^5$. 
Despite the simplicity of the model, we show that the transition belongs to the same universality class as fracton models with identical critical exponents. Furthermore, we study the Krylov-sector-restricted dynamical structure factor of the model, and find that charge transport is diffusive in the thermal phase, which is to be contrasted with previous results without resolving the Krylov sectors~\cite{PhysRevLett.127.230602}. Our results provide a tractable minimal model of disorder-free dynamical phase transition that can be readily tested in state-of-the-art quantum platforms using controlled-unitary gates.

\indent\textit{Model.-} We study a one-dimensional system of $N$ hardcore bosonic particles with nearest-neighbor hopping on $L$ lattice sites. Each site $i$ can host $n_i=0$ or 1 particle, and one can equivalently consider a qubit or spin-1/2 system where the computational or spin-$z$ basis configurations correspond to particle occupations. We use open boundary condition unless otherwise specified. 
We further impose an East-like kinetic constraint on the dynamics: a particle can hop to the right only when there is an occupied site within a distance $r$ to its left (i.e., an occupied site can mobilize nearby particles to its ``east"), as illustrated in Fig.~\ref{fig:east}(a) for $r=1$ and $r=2$. Such a constraint is inspired by the East model~\cite{faggionato2012east}, or more generally, facilitated spin models in classical glasses~\cite{ritort2003glassy}, where spin flips are facilitated by an adjacent spin along a particular orientation. Recently, its quantum versions (without particle number or $S^z$ conservation) have been proposed as candidates for slow thermalization and localization without disorder~\cite{PhysRevB.92.100305, PhysRevX.10.021051, bertini2023localised, PRXQuantum.3.020346}. Introducing an additional U(1) particle number conservation allows for the study of HSF~\cite{brighi2022hilbert} as well as transport properties of the conserved charge~\cite{PhysRevLett.127.230602}. While one can construct Hamiltonians
generating the constrained dynamics in Fig.~\ref{fig:east}(a), the essential physics of HSF and the freezing transition does not require a time-independent Hamiltonian. Therefore, we consider more generally dynamics generated by classical Markovian circuits, which is equivalent to quantum automaton circuits when starting from a single particle configuration with fixed occupation numbers on each site. 

The circuit consists of consecutive layers of $(r+2)$-site gates. Take $r=1$ as an example: local three-site gates implement the moves: $\bullet \circ \bullet \leftrightarrow \bullet \bullet \circ$, where $\bullet$ and $\circ$ denote an occupied and empty site, respectively. As a result of the kinetic constraint, not all particle configurations belonging to the same charge sector can be connected to one another under the dynamics, and hence the Hilbert space further fractures into Krylov subsectors. To see this, notice that according to Fig.~\ref{fig:east}(a), the position of the leftmost particle is conserved under the dynamics, and thus configurations with distinct leftmost particle positions cannot be connected by the dynamical moves even if they have the same particle number.

\textit{Freezing transition.-} Intuitively, it is easy to understand why the average particle density can affect the degree of fragmentation in this model. At low fillings, particles in the system are well isolated from one another, and it is very unlikely to find an occupied site to the left of a particle to trigger hopping. Thus, most of the particles are frozen and the Hilbert space is strongly fragmented. At high fillings, it is almost always possible to find a nearby occupied site, and the constraint essentially becomes ineffective. Therefore, we expect the structure of the Hilbert space to change qualitatively as the density is varied. 

To quantify the degree of HSF, it is useful to consider the ratio of the dimension of the largest Krylov sector and that of the entire symmetry sector $D_{\rm max}/D_{\rm sum}$. While the total size of a symmetry sector with $N$ particles is simply $D_{\rm sum} = {L \choose N}$, an analytic expression for $D_{\rm max}$ is usually hard, and numerically enumerating all configurations within a Krylov sector is only possible for very small system sizes. In fact, it is in general difficult to even identify the largest Krylov sector within each symmetry sector. However, for the East models, we are able to develop a simple algorithm for computing $D_{\rm max}$ recursively, and even analytic solutions in certain cases.

To begin with, it is easy to show that within a symmetry sector of $N$ particles, the largest Krylov sector is generated from the following root configuration:
\begin{equation}
\underbrace{\bullet \bullet \bullet \cdots \bullet}_N \underbrace{\circ \circ \cdots \circ}_{L-N},
\label{eq:root}
\end{equation}
i.e., a domain wall configuration with all $N$ particles occupying the first $N$ sites from the left. The reason is that this sector has only one frozen particle which is the leftmost particle, and hence one active block. If there are more than one active blocks separated by frozen regions, one can always form a different Krylov sector by concatenating the active blocks and moving all frozen regions to the right. The resulting Krylov sector is necessarily larger than the original one, and hence the largest Krylov sector is generated by particle configuration~(\ref{eq:root}).

\begin{figure}[!t]
\includegraphics[width=0.5\textwidth]{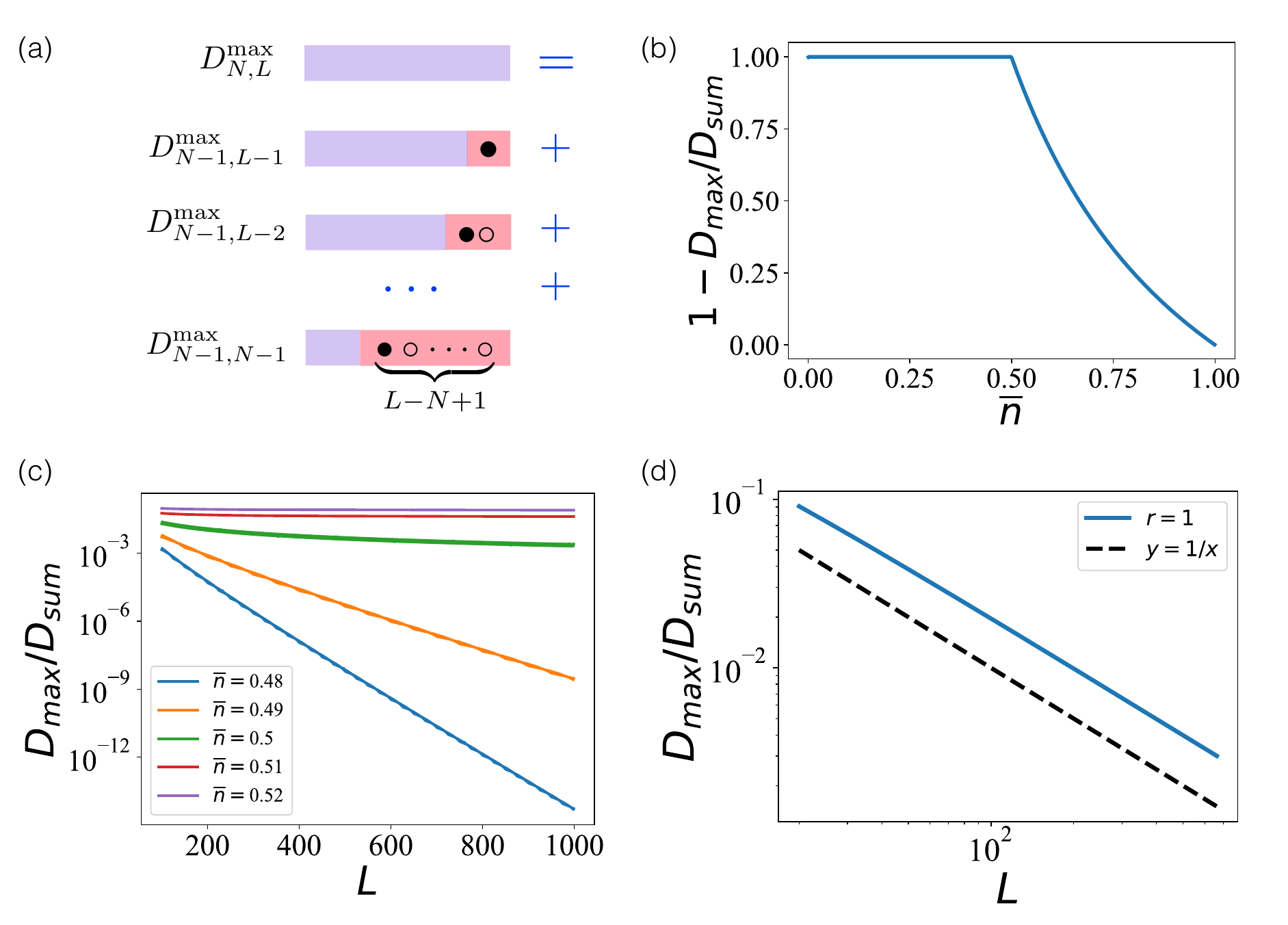}
\caption{(a) Recursive algorithm for computing the size of the largest Krylov subspace $D^{\rm max}_{N,L}$ for $N$ particles on $L$ sites. The position of the rightmost particle in the red shaded region is held fixed in each term. (b) The ratio $1-D_{\rm max}/D_{\rm sum}$ as a function of the particle density for system size $L=1000$, which shows a phase transition at $n_c=0.5$. (c) Scaling of the ratio $D_{\rm max}/D_{\rm sum}$ with $L$ below and above the critical filling. For $n>n_c$, the ratio saturates to order one as $L$ increases, indicating weak fragmentation. For $n<n_c$, the ratio decays exponentially with $L$, indicating strong fragmentation. (d) At the critical point, the fraction of the largest Krylov sector shows a power-law decay with system size: $D_{\rm max}/D_{\rm sum} \sim L^{-1}$. }
\label{fig:D_ratio} 
\end{figure}

After identifying the largest Krylov sector, we have yet to compute its size. This can be done recursively in the East model, as illustrated in Fig.~\ref{fig:D_ratio}(a). First of all, starting from the root configuration~(\ref{eq:root}), the longest distance that the particles can spread is given by $L_{\rm max}=(r+1)N - r$, corresponding to the most dilute particle configuration:
\begin{equation}
\underbrace{\bullet \circ \cdots \circ}_{r+1} \underbrace{\bullet \circ \cdots \circ}_{r+1} \bullet \cdots. 
\end{equation}
Denote the dimension of the largest Krylov sector with $N$ particles on $L$ sites as $D^{\rm max}_{N,L}$. Apparently, we have $D^{\rm max}_{N,L} = D^{\rm max}_{N, L_{\rm max}}$ for $L > L_{\rm max}$. For $L\leq L_{\rm max}$, one can obtain $D^{\max}_{N,L}$ from the dimensions of Krylov sectors of the same type [i.e. those that are generated from the root configuration~(\ref{eq:root})] with $(N-1)$ particles on $L-1$, $L-2$, $\ldots$, $N-1$ lattice sites, which corresponds to fixing the rightmost particle at all possible positions [see Fig.~\ref{fig:D_ratio}(a)]:
\begin{equation}
D^{\rm max}_{N,L} = D^{\rm max}_{N-1, L-1} + D^{\rm max}_{N-1, L-2} + \cdots + D^{\rm max}_{N-1, N-1}.
\end{equation}
To summarize, we have the following recursion relation:
\begin{equation}
    D^{\rm max}_{N,L}=
    \begin{cases}
        D^{\rm max}_{N,L_{\rm max}},   & L > L_{\rm max} \\
                                                        \\
        \sum\limits_{i=N-1}^{L-1} D^{\rm max}_{N-1, i},     &L \leq L_{\rm max}
    \end{cases}.
\end{equation}
Carrying out the above recursion relation up to system size $L$ with particle numbers $N\leq L$ requires only $\mathcal{O}(L^2)$ operations, which allows us to efficiently compute the size of the largest Krylov sector up to $L\sim 10^3$ and obtain clear signatures of a phase transition.

In Fig.~\ref{fig:D_ratio}(b)-(d), we show numerical results for $r=1$ using the recursive algorithm described above. Fig.~\ref{fig:D_ratio}(b) clearly shows a transition in the ratio $D_{\rm max}/D_{\rm sum}$ as the average density is varied. For $n<0.5$, $D_{\rm max}$ constitutes a small fraction of the entire symmetry sector, indicating strong fragmentation. For $n>0.5$, the ratio approaches order one, indicating weak fragmentation. We further consider the scaling of the ratio $D_{\rm max}/D_{\rm sum}$ with $L$ below and above the critical filling, as shown in Fig.~\ref{fig:D_ratio}(c). In the weakly fragmented phase, this ratio saturates to a constant of order unity as $L$ increases. In the strongly fragmented phase, the ratio decays exponentially with $L$, which implies that even the size of the largest Krylov subsector is vanishingly small compared with the full symmetry sector in the thermodynamic limit. At the critical point, we find that $D_{\rm max}/D_{\rm sum}$ exhibits a power-law decay with system size: $D_{\rm max}/D_{\rm sum} \sim 1/L$, as shown in Fig.~\ref{fig:D_ratio}(d). In the Supplemental Material (SM)~\cite{SM}, we prove that for $r=1$, $D_{\rm max}$ at the critical point (with $L=2N$) is precisely given by the $N$-th Catalan number:
\begin{equation}
D^{\rm max}_{N, 2N}= C_N \equiv \frac{1}{N+1} {2N \choose N} = \frac{1}{N+1} D_{\rm sum}.
\end{equation}
Thus, the ratio $D_{\rm max}/D_{\rm sum} \sim L^{-1}$, which explains our numerical finding in Fig.~\ref{fig:D_ratio}(d). 

The qualitative change in the structure of the Hilbert space as diagnosed by the ratio $D_{\rm max}/D_{\rm sum}$ has a direct consequence on the dynamics of the system, starting from an initial state at a given filling $n$. We consider the average density of frozen sites $\langle n_F\rangle$, defined as the fraction of sites whose occupation numbers remain unchanged under the circuit dynamics at infinite times~\cite{PhysRevB.101.214205}. This quantity is averaged over all initial states within the same charge sector, and serves as an order parameter for the transition. Notice that a site is frozen if its occupation is the same in all configurations within the same Krylov sector, and hence $\langle n_F \rangle$ is closely related to the connectivity of the Hilbert space. This order parameter, however, is in general hard to compute. Usually one has to either enumerate all Krylov sectors for small system sizes (infinite time, finite size regime), or simulate the dynamics for large systems at early times (infinite system, finite time regime)~\cite{PhysRevB.101.214205}. Fortunately, the East model allows us to access both infinite time and infinite system limit simultaneously via an efficient way of simulating the dynamics of growing thermal bubbles. We defer the detailed algorithm to the next section, and show our numerical result of $\langle n_F \rangle$ in Fig.~\ref{fig:east}(b), which we obtain from sampling $10^3$ different configurations of size $L=10^3$ at a given filling and compute the average $n_F$. The result clearly shows that $\langle n_F \rangle$ is zero for $n>0.5$ (thermal phase) and becomes nonzero for $n<0.5$ (frozen phase). Furthermore, we find that near the transition, $\langle n_F \rangle \sim (n_c-n)^\beta$ with $\beta =1$.

The critical density $n_c$ turns out to be quite straightforward to compute for the East model. Consider the largest Krylov sector~(\ref{eq:root}). Since the longest distance that the particles can spread is $L_{\rm max}$, for $L<L_{\rm max}$, all sites are necessarily active; for $L>L_{\rm max}$, there will be sites on the right end that cannot be reached by particles, and a non-zero fraction of frozen sites will emerge. Therefore, the critical density is given by
\begin{equation}
n_c = \frac{N}{L_{\rm max}} \xrightarrow{L \rightarrow \infty} \frac{1}{1+r}.
\label{eq:critical}
\end{equation}
For $r=1$, this is in agreement with our numerical results. In the SM~\cite{SM}, we provide numerical results for $r=2$ which also shows perfect agreement with Eq.~(\ref{eq:critical}).

\begin{figure}[!t]
\includegraphics[width=0.48\textwidth]{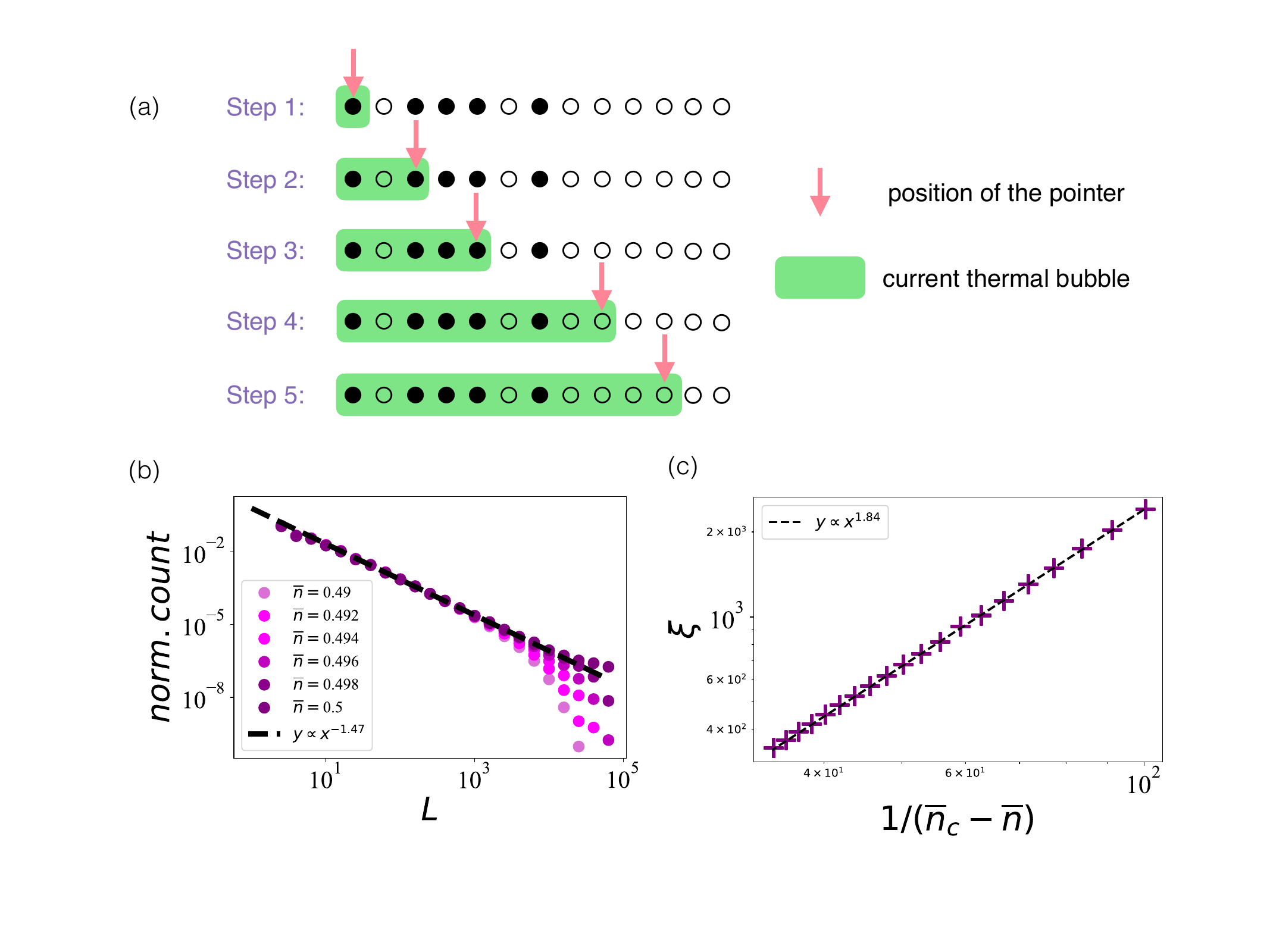}
\caption{(a) Illustration of the procedure for finding the maximal size that a thermal region can grow into. The red arrow is a pointer that marks the rightmost point of the growing thermal bubble, and the green region denotes sites that have already been absorbed into the bubble at the current step. (b) Distribution of the ultimate sizes of the thermal region at infinite times near $n_c$. Numerical results are obtained by carrying out the procedure depicted in (a) for $10^3$ samples of initial configurations of system size $L=10^5$. The distribution exhibits a power law decay $P(l)\sim l^{-3/2}$. (c) The correlation length diverges as $\xi \sim (n_c-n)^{-\nu}$ with $\nu \approx 2$.} 
\label{fig:bubble} 
\end{figure}

\textit{Thermal inclusion.-} For $n \lesssim n_c$, a large sample of the system typically contains local thermal regions with $n>n_c$, as well as frozen regions with $n<n_c$. Under time evolution, excess particles in the thermal region will propagate into nearby frozen regions and absorb them into a larger thermal region. Of course, this process will decrease the charge density of the thermal region, and hence the growth of a thermal bubble stops once its average filling decreases to $n_c$. We study this thermal inclusion process for the East model near the critical point.

We find an efficient way of figuring out the maximal size that an initial thermal seed can grow into at infinite times for the East model. For a random initial particle configuration, we use a pointer that starts from the leftmost site and moves towards the right, until we reach the first particle and start counting the size of the current thermal bubble. We then add more sites that can be absorbed into the bubble by moving the pointer further to the right according to the following rule. We compute the total number of particles $N$ currently in the bubble, and move the pointer to site $(r+1)N+1=L_{\rm max}+(r+1)$ counting from the leftmost site in the bubble, which is the farthest site that the current bubble can affect. The above step is repeated until no additional particle is encountered between two consecutive moves of the pointer, indicating that the thermal bubble cannot grow any further to the right at this point. 
We record the length of this thermal region, and start over by moving our pointer to the right until we reach a new particle, and start counting the size of the next thermal region. The procedure continues until we reach the rightmost site of the system. We give a concrete example of this algorithm in Fig.~\ref{fig:bubble}(a). Apparently, this procedure requires only $\mathcal{O}(L)$ operations, and can be carried out for extremely large system sizes.

In Fig.~\ref{fig:bubble}(b), we find that the ultimate sizes of the thermal regions follow a power-law distribution near the critical point $P(l)\sim l^{-3/2}$ for $l<\xi$, where $\xi$ is identified as the correlation length. We can further extract $\xi$ from the moments of $P(l)$: $\xi = \langle l^2 \rangle / \langle l \rangle$. We find that the correlation length diverges as $\xi \sim (n_c-n)^{-\nu}$ with $\nu \approx 2$ near the critical point. Interestingly, the critical exponents we obtained by explicitly growing all thermal bubbles microscopically is in perfect agreement with a simplified effective model constructed for the fracton model~\cite{PhysRevB.101.214205}. In the SM, we further show that these exponents remain the same for $r=2$. We are thus led to conclude that the universality class of the freezing transition is largely independent of the microscopic details of the model, as long as it is driven by charge density, and the underlying physics is captured by the growth of local thermal bubbles with $n>n_c$ until they self-tune to the critical density.
We now have an explicit example where the validity of the effective model proposed in Ref.~\cite{PhysRevB.101.214205} is confirmed via exact numerical simulations of the microscopics.

\begin{figure}[!t]
\includegraphics[width=0.3\textwidth]{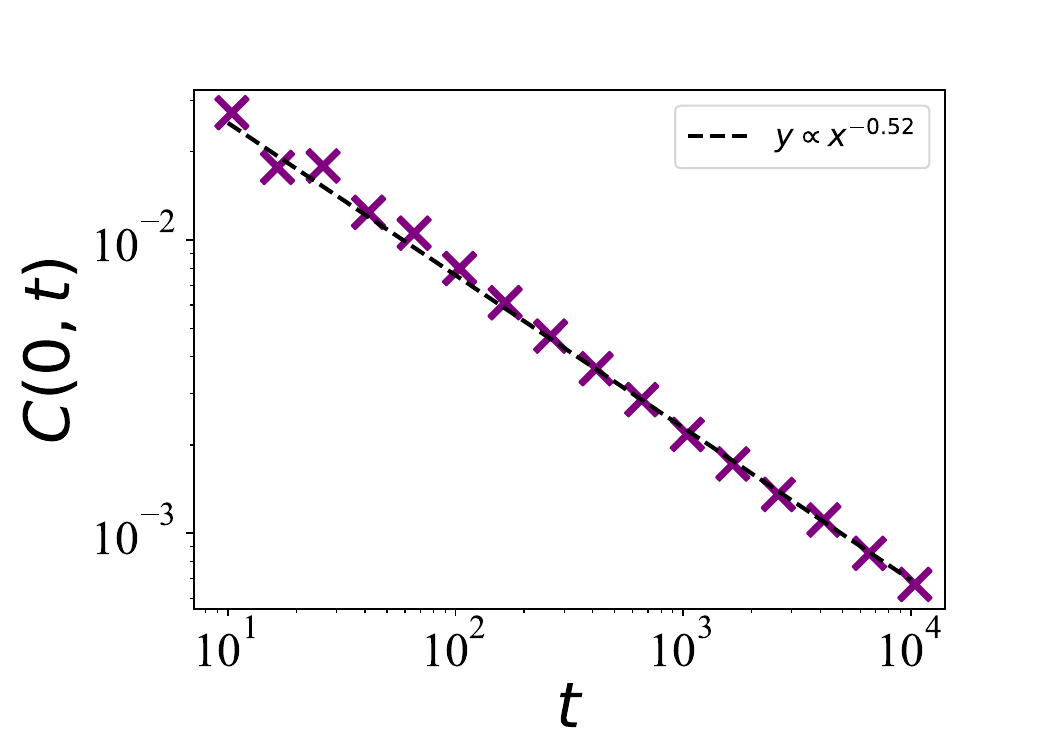}
\caption{The autocorrelation function restricted to a specific charge sector with $n>n_c$ decays as $C(0,t)\sim t^{-1/2}$, indicating diffusive transport $z=2$ in the thermal phase. Results are obtained by sampling $10^{3}$ initial states of size $L=1000$ with an average particle density $n=0.8>n_c$.}
\label{fig:transport} 
\end{figure}

\textit{Krylov-sector-restricted transport.-} Finally, we study transport properties of the particle-conserving East model in the thermal phase. We compute the infinite-temperature autocorrelation function of the charge density on a given site $C(0,t) \equiv {\rm tr}[n_i(t) n_i(0)]/D$, where the trace is restricted to configurations within a given charge sector, and $D$ denotes the size of this sector. We use periodic boundary condition and average over all sites. In Fig.~\ref{fig:transport}, we find that the autocorrelation function at long times decay as $C(0,t)\sim t^{-1/2}$, consistent with diffusive transport with $z=2$. This result is easy to understand: since the kinetic contraint is ineffective in the thermal phase, particles hop around as in an unconstrained U(1) symmetric system, and hence charge transport obeys diffusion. Notice, however, that this result is in sharp contrast to previous studies where this correlator is averaged over \textit{all} symmetries sectors, which leads to a diverging dynamical exponent at late times~\cite{PhysRevLett.127.230602}. Our results clarify the origin of this distinction: the dynamics within each U(1) sector actually undergo a phase transition, and hence it is crucial to study the Krylov-sector resolved transport properties. Recently, the existence of diffusive Krylov sectors in subdiffusive dipole-conserving systems has also been demonstrated~\cite{ogunnaike2023unifying}.

\textit{Summary and outlook.-} We study a particle-conserving East model in which particle hoppings are facilitated by the presence of other particles to its left. We find that the structure of the Hilbert space and the dynamical properties exhibit a sharp transition as the average particle density is varied, going from weakly fragmented and thermal at high fillings $n>n_c$ to strongly fragmented and frozen at low fillings $n<n_c$. The special feature of the model allows for both analytic solutions and efficient numerical simulations which are combined to characterize the universal properties at the transition. Despite its simplicity, we find that the transition belongs to the same universality class as in dipole-conserving fracton models, where the microscopics are much more complicated. Our results thus provide a tractable minimal model for filling-induced freezing transitions in quantum many-body systems. The East-like constraint can be implemented via controlled-unitary gates, and hence the physics explored in this work can be readily tested in state-of-the-art quantum platforms such as trapped ions and superconducting qubits, using random circuit evolutions. 

\textit{Acknowledgments.-} We thank Jingwu Tang for helpful discussions on Catalan number. Z.-C.Y. is supported by a startup fund at Peking University. Numerical simulations were performed on High-performance Computing Platform of Peking University.

\bibliography{reference}

\newpage
\onecolumngrid
\appendix

\subsection*{Supplemental Material for ``Freezing transition in particle-conserving East model"}

\section{Analytic expressions for the size of the largest Krylov sector for $r=1$}

In this section, we give analytic expressions for the sizes of the largest Krylov sectors for $r=1$. We start from the critical filling $n=0.5$, or $L=2N$. The largest Krylov sector is generated from the root configuration:
\begin{equation}
\underbrace{\bullet \bullet \bullet \cdots \bullet}_N \underbrace{\circ \circ \cdots \circ}_{N}.
\end{equation}
Since this sector is fully connected, configurations belonging to this sector cannot have any frozen site in the bulk that separates the entire system into disconnected regions. Therefore, the allowed configurations must satisfy the following condition: for any bipartitioning of the system into $A=[1,k]$ and $\overline{A}=[k+1,L]$, there cannot be more empty sites than occupied sites within region $A$. For example, $\bullet \circ \bullet \circ \circ \bullet \cdots$ cannot reside within this subsector, and is hence forbidden.

\begin{figure}[!b]
\includegraphics[width=0.6\textwidth]{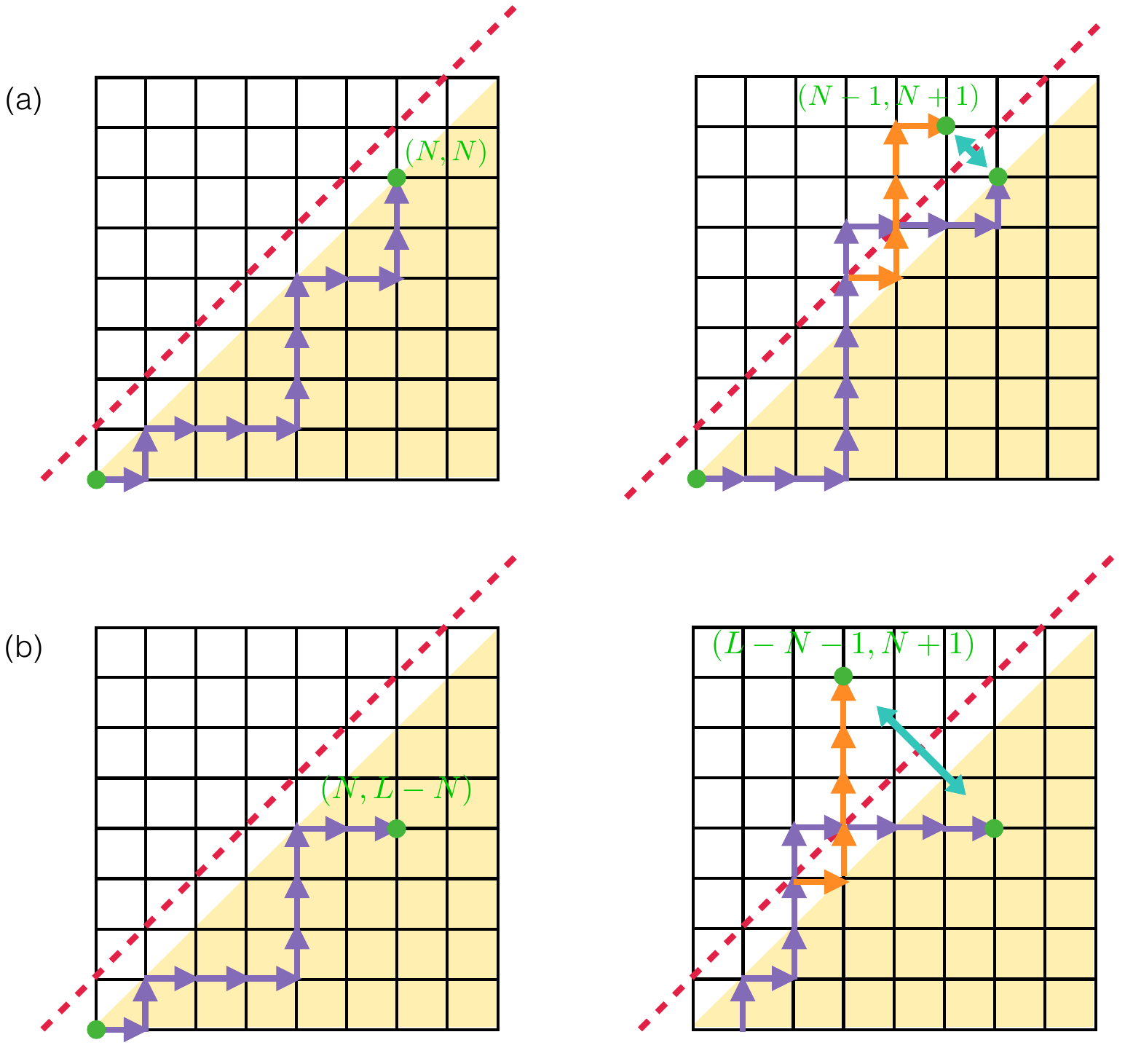}
\caption{Computing the size of the largest Krylov sector by mapping to the combinatorial problem of counting the number of allowed monotonic paths on a lattice. (a) $n=0.5$; (b) $n>0.5$. The paths connect the two green dots on the lattice, and are restricted within the yellow shaded region, i.e., they cannot touch or cross the red dashed line $y=x+1$ (left pannel). We can count the number of disallowed paths by mapping such paths to those connecting the origin and the mirror-reflected point about the red line (orange paths in the right pannel). }
\label{fig:dimension} 
\end{figure}

Counting the total number of configurations satisfying the abovementioned condition is a well-known problem in combinatorics. The problem is equivalent to counting the number of Dyck words of length $2N$, with occupied and empty sites corresponding to two different alphabets. The solution is given by the $N$-th Catalan number:
\begin{equation}
C_N \equiv \frac{1}{N+1} {2N \choose N}.
\end{equation}

To generalize the above result to $n>0.5$, it is useful to first introduce an alternative interpretation of the combinatorial problem. Consider a square lattice grid as depicted in Fig.~\ref{fig:dimension}. For a configuration with $N$ particles and $N$ holes, we start from the origin of the lattice, and draw a horizontal arrow $\rightarrow$ each time we see a particle, and a vertical arrow $\uparrow$ for each hole. We end up with a \textit{monotonic} path connecting the origin $(0,0)$ and site $(N,N)$ on the lattice. Here, monotonicity simply means that there is no left or down pointing arrow, and the path contains precisely $2N$ steps.
However, due to the constraint that there cannot be more empty sites than occupied sites for any contiguous subregions including the leftmost site, the allowed paths can only stay in the yellow shaded region depicted in Fig.~\ref{fig:dimension}. In particular, they cannot touch or cross the red dashed line $y=x+1$. To count the total number of allowed paths, we need to substract from all paths connecting the two points those that are disallowed. There is a simple way of counting the number of disallowed paths. As we illustrate in Fig.~\ref{fig:dimension}(a), suppose a path touches or crosses the red dashed line. We do a reflection of the path starting from the crossing point about the red line [orange path in Fig.~\ref{fig:dimension}(a)]. The reflected path now connects the origin to the reflected site $(N-1, N+1)$. It is easy to see that, disallowed paths have a one-to-one correspondence to all possible paths connecting the origin and the reflected site $(N-1, N+1)$. Hence, we interpret the Catalan number as the subtraction of disallowed paths from all possible paths:
\begin{equation}
C_N = {2N \choose N} - {2N \choose N-1}.
\end{equation}

Using this interpretation, it is straightforward to obtain an analytic expression for the size of the largest Krylov sector for $n>0.5$. In this case, we have $N$ particles and $L-N<N$ holes, and configurations belonging to this Krylov sector can be mapped to all paths connecting the origin and site $(N, L-N)$ restricted in the yellow shaded region, as depicted in Fig.~\ref{fig:dimension}(b). Using the same trick of mapping disallowed paths to paths connecting the origin and the mirror-reflected point, we obtain the total number of allowed paths:
\begin{equation}
D^{\rm max}_{N,L} = {L \choose N} - {L \choose N+1}.
\end{equation}

\section{Numerical results for $r=2$}

In this section, we present additional numerical results for the East model with range $r=2$. We will see that the essential physics discussed in the main text is independent of the range $r$.

\begin{figure}[!t]
\includegraphics[width=0.8\textwidth]{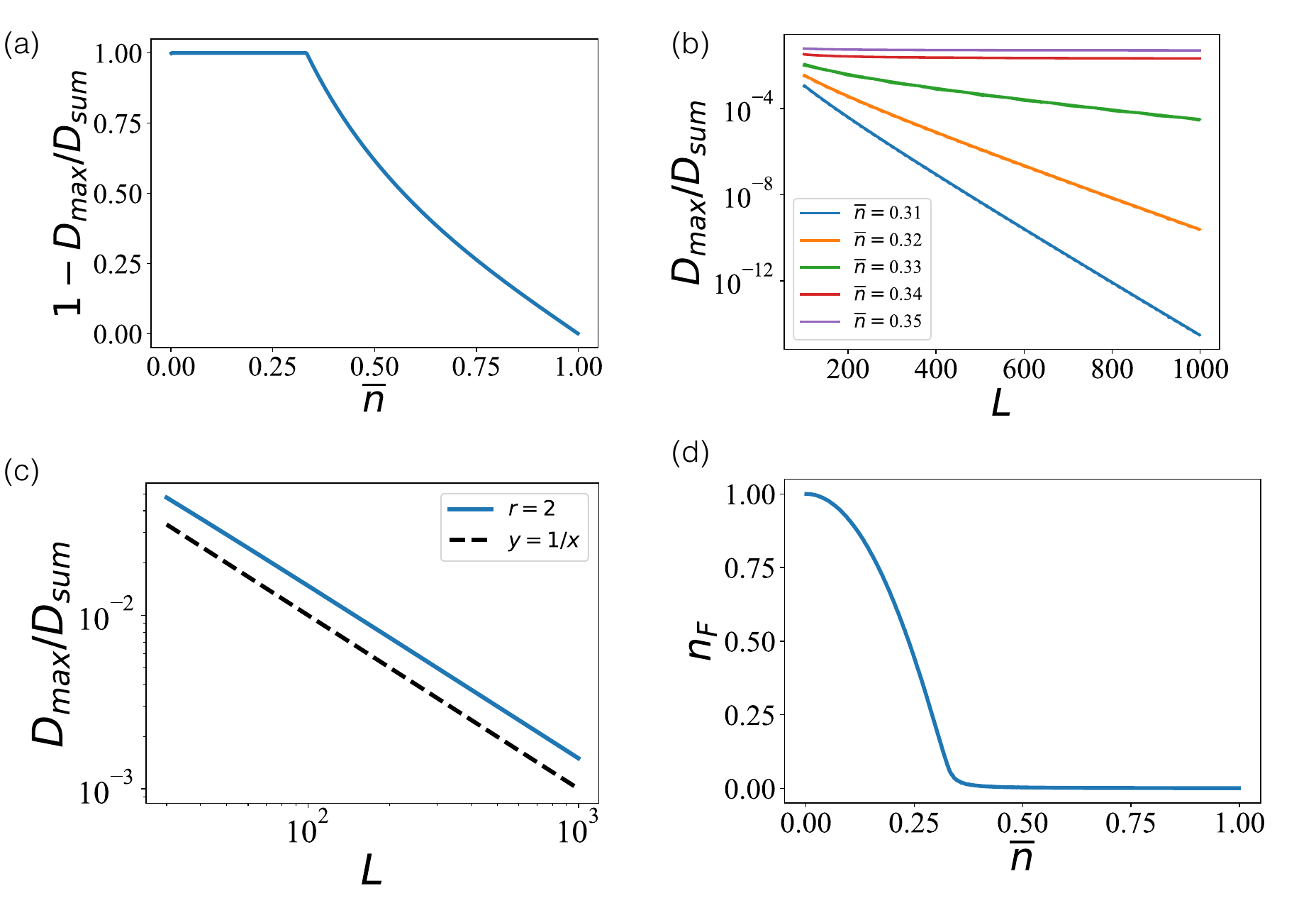}
\caption{Freezing transition for the $r=2$ particle-conserving East model. (a) The ratio $1-D_{\rm max}/D_{\rm sum}$ as a function of the particle density for system size $L=1000$, which shows a phase transition at $n_c=1/3$. (b) Scaling of the ratio $D_{\rm max}/D_{\rm sum}$ with $L$ below and above the critical filling. For $n>n_c$, the ratio saturates to order one as $L$ increases, indicating weak fragmentation. For $n<n_c$, the ratio decays exponentially with $L$, indicating strong fragmentation. (c) At the critical point, the fraction of the largest Krylov sector shows a power-law decay with system size: $D_{\rm max}/D_{\rm sum} \sim L^{-1}$. (d) The average fraction of frozen sites $\langle n_F \rangle$ sampled over $10^3$ different configurations of size $L=1000$. Near $n_c$, we find $\langle n_F\rangle \sim (n_c-n)^\beta$ with $\beta=1$.}
\label{fig:D_ratior2} 
\end{figure}

We start by showing our diagnostics for the phase transition in Fig.~\ref{fig:D_ratior2}. Notice that in this case, we do not have analytic expressions for the size of the largest Krylov sector as in the case of $r=1$. So we implement the recursive algorithm outlined in the main text, which works independent of the range $r$.
The ratio $D_{\rm max}/D_{\rm sum}$ exhibits a qualitative change at the critial density $n_c \approx 0.33$ [Fig.~\ref{fig:D_ratior2}(a)]. For $n<n_c$, the largest Krylov sector constitute a vanishingly small fraction of the full symmetry sector in the thermodynamic limit, indicative of strong fragmentation. The ratio decays exponentially with $L$ upon increasing system size in this regime [Fig.~\ref{fig:D_ratior2}(b)]. For $n>n_c$, the ratio approaches order one, indicating weak fragmentation, and the system thermalizes with high probability from a random initial state. At the critical point, $D_{\rm max}/D_{\rm sum}$ again shows a power-law decay with system size as $L^{-1}$. We can similarly consider the fraction of frozen sites averaged over all configurations in a symmetry sector as a function of the filling, which serves as an order parameter for the transition. This order parameter changes from zero to nonzero at the critical $n_c$, as shown in Fig.~\ref{fig:D_ratior2}(d). Notice that the position of the critical point is again in excellent agreement with the general expression $n_c=\frac{1}{r+1}$, which is equal to $1/3$ for $r=2$.

\begin{figure}[!t]
\includegraphics[width=0.8\textwidth]{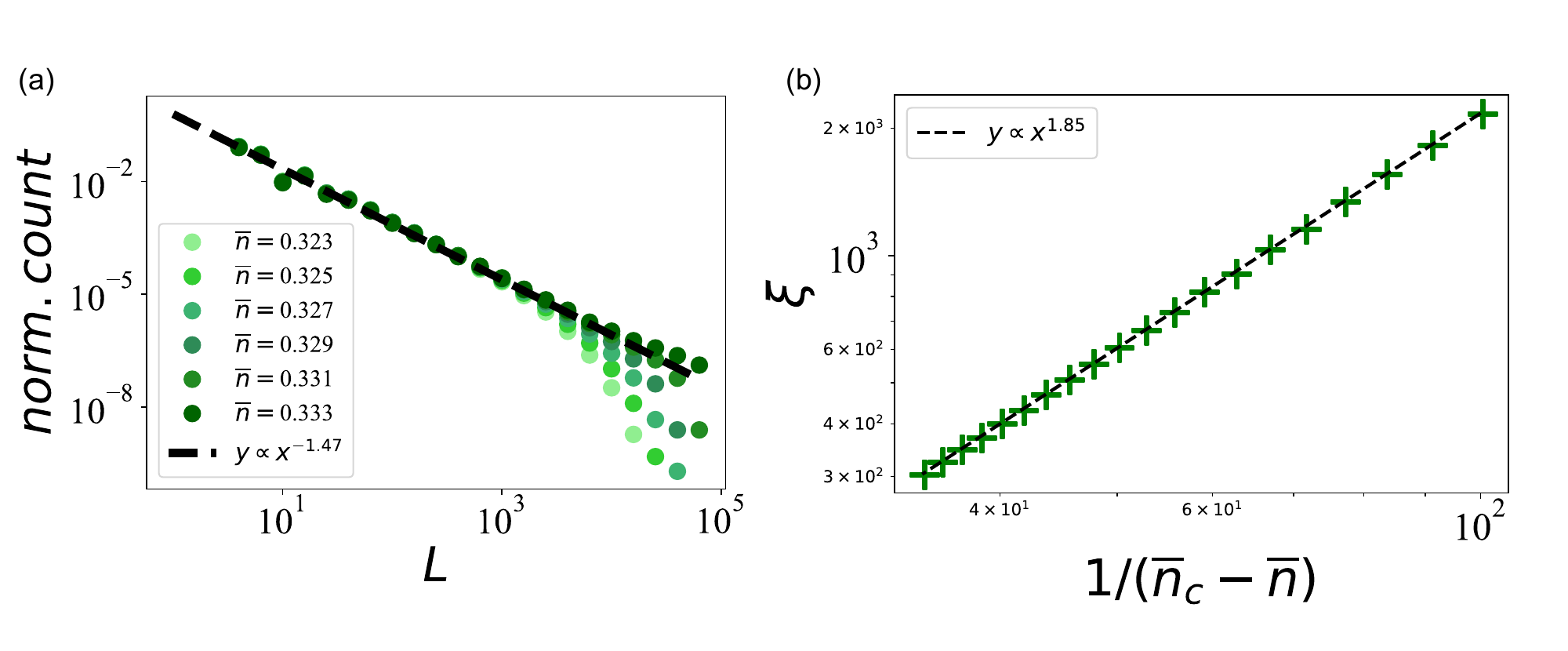}
\caption{(a) Distribution of the ultimate sizes of the thermal region at infinite times near $n_c$. Numerical results are obtained by carrying out the procedure explained in the main text for $10^3$ samples of initial configurations of system size $L=10^5$. The distribution exhibits a power law decay $P(l)\sim l^{-3/2}$. (b) The correlation length diverges as $\xi \sim (n_c-n)^{-\nu}$ with $\nu \approx 2$. }
\label{fig:bubble_r2} 
\end{figure}

We also consider the process of thermal inclusion in this case, for which numerical results are summarized in Fig.~\ref{fig:bubble_r2}. We find that the distribution of the sizes of the thermal bubble again obeys $P(l)\sim l^{-3/2}$ for $l<\xi$, and the correlation length itself diverges as $\xi \sim (n_c-n)^{-\nu}$ with $\nu \approx 2$.

\begin{figure}[!t]
\includegraphics[width=0.5\textwidth]{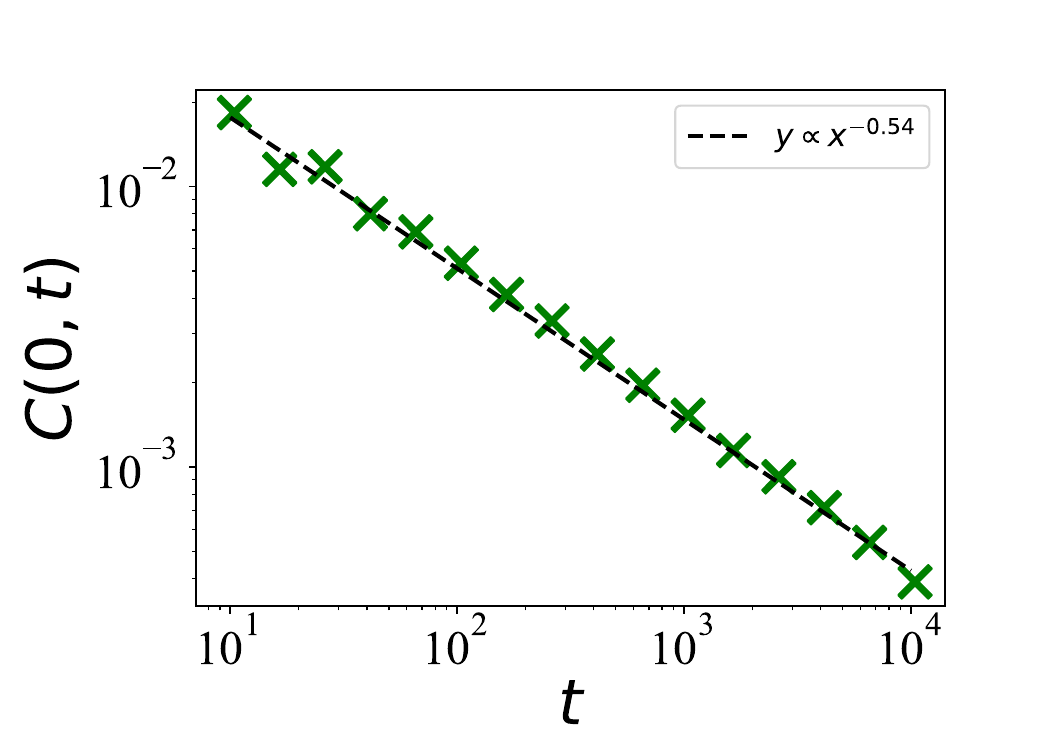}
\caption{The autocorrelation function restricted to a specific charge sector with $n>n_c$ decays as $C(0,t)\sim t^{-1/2}$, indicating diffusive transport $z=2$ in the thermal phase. Results are obtained by sampling $10^{3}$ initial states of size $L=1000$ with an average particle density $n=0.8$.}
\label{fig:transport_r2} 
\end{figure}

Finally, we confirm that charge transport is diffusive in the thermal phase by computing the autocorrelation function restricted to a specific symmetry sector, as shown in Fig.~\ref{fig:transport_r2}.

\end{document}